\documentclass[JIP]{ipsj}

\usepackage{cite}
\usepackage{physics}
\usepackage{graphicx}
\usepackage[font=small]{caption}
\usepackage{subcaption}

\usepackage{tikz}
\usetikzlibrary{calc,matrix,positioning}

\usepackage{url}

\begin{document}

\title{Performance Evaluation of Parallel Sortings on the Supercomputer Fugaku}

\affiliate{CU}{Department of Applied and Cognitive Informatics, Division of Mathematics and Informatics, Graduate School of Science and Engineering, Chiba University}
\affiliate{CUDTEC}{Digital Transformation Enhancement Council, Chiba University}

\author{Tomoyuki Tokuue}{CU}[t.tokuue@chiba-u.jp]
\author{Tomoaki Ishiyama}{CUDTEC}[ishiyama@chiba-u.jp]

\begin{abstract}
Sorting is one of the most basic algorithms, and developing highly parallel sorting programs is becoming increasingly important in high-performance computing because the number of CPU cores per node in modern supercomputers tends to increase. In this study, we have implemented two multi-threaded sorting algorithms based on samplesort and compared their performance on the supercomputer Fugaku. The first algorithm divides an input sequence into multiple blocks, sorts each block, and then selects pivots by sampling from each block at regular intervals. Each block is then partitioned using the pivots, and partitions in different blocks are merged into a single sorted sequence. The second algorithm differs from the first one in only selecting pivots, where the binary search is used to select pivots such that the number of elements in each partition is equal. We compare the performance of the two algorithms with different sequential sorting and multiway merging algorithms. We demonstrate that the second algorithm with BlockQuicksort (a quicksort accelerated by reducing conditional branches) for sequential sorting and the selection tree for merging shows consistently high speed and high parallel efficiency for various input data types and data sizes.
\end{abstract}


\maketitle

\section{Introduction}
Sorting is one of the most fundamental algorithms, and fast sorting programs are required in tons of applications. Therefore the optimization and parallelization of sorting in shared/distributed memory environments have always been a research subject~\cite{Pquick,Solomonik2010,PBBS,RegionsSort}. Developing highly parallel sorting programs is becoming increasingly important in high-performance computing because the number of computational cores per node in modern supercomputers tends to increase. For example, the supercomputer Fugaku, which is a flagship supercomputer in Japan, consists of 48 computational cores per node. Even though this number is expected to increase further in future supercomputers, few sorting programs with high parallel efficiency for many threads have been reported.

In this study, we develop comparison-based sorting programs that run at high speed and are efficiently parallelized on Fugaku. We implement two multi-threaded sorting algorithms based on samplesort~\cite{Samplesort} and compare the performance of each algorithm on Fugaku for various input data types and data sizes. We also compare sequential sorting and multiway merging algorithms used in samplesort and find the best combination. For the sequential sorting algorithm, we adopt quicksort~\cite{quicksort} and its variants (e.g., \cite{blockQsort,pdqsort}). Quicksort is generally regarded as a faster sorting algorithm than the others. For multiway merging, we compare the methods with and without data structures. We adopt the binary heap and the selection tree~\cite{Knuth1998} as data structures for efficient merging. The source code is publicly available at \url{https://github.com/tmtoku/parallel_sortings}.

This paper is organized as follows. In Section~\ref{sec:method}, we describe parallel sorting algorithms and introduce sequential sorting and multiway merging algorithms used in this study. We evaluate the performance of the parallel sorting algorithms on the supercomputer Fugaku in Section~\ref{sec:evaluations} and summarize the results in Section~\ref{sec:conclusion}.

\section{Parallel samplesort}\label{sec:method}
Samplesort~\cite{Samplesort}, which is widely used in parallel environments, picks multiple pivots by sampling, partitions an input sequence, and sorts each partition.

Many parallel samplesort~\cite{Shi1992,Siebert2011,MCSTL} for an $N$-elements sequence $A$ is done in the following four steps.
\begin{enumerate}
        \item Sorting each block\label{procedure:blockSort}
        \item Pivots selection\label{procedure:pivotSelection}
        \item Partitioning\label{procedure:partitioning}
        \item Multiway merging of partitions\label{procedure:merge}
\end{enumerate}
First, it divides $A$ into $n_\text{B}$ blocks and sorts each block sequentially (step~\ref{procedure:blockSort}). A block is a contiguous subsequence of length $\lceil N/n_\text{B}\rceil$. Then, it selects $n_\text{P}-1$ pivots $P_1,\,\ldots,\,P_{n_\text{P}-1}$ in step~\ref{procedure:pivotSelection} and rearranges each block into $n_\text{P}$ partitions in step~\ref{procedure:partitioning}. With $P_0=-\infty$ and $P_{n_\text{P}}=\infty$ (assuming sorting in the ascending order), the $k$th partition $\qty(0\leq k<n_\text{P})$ is a contiguous subsequence consisting of elements $x$ of $P_k<x\leq P_{k+1}$. Finally, step~\ref{procedure:merge} merges the $k$th partition in all blocks and puts them in order. Step~\ref{procedure:blockSort} and step~\ref{procedure:partitioning} can be performed in parallel for different blocks, and step~\ref{procedure:pivotSelection} for different pivots. Multiway merging of each partition in step~\ref{procedure:merge} can be executed in parallel, but we have to sequentially calculate the offset of each merged partition from the beginning of the output sequence.

\begin{figure}
    \centering
    {\sffamily
    \begin{tikzpicture}[MyStyle/.style={draw, minimum width=2em, minimum height=2em, outer sep=0pt}]
    \matrix (A1) [matrix of math nodes, nodes={MyStyle, anchor=center}, column sep=-\pgflinewidth]{5 & 2 & 8\\};
    \matrix[right=.5em of A1] (B1) [matrix of math nodes, nodes={MyStyle, anchor=center}, column sep=-\pgflinewidth]{4 & 1 & 7\\};
    \matrix (A2) [matrix of math nodes, nodes={MyStyle, anchor=center}, column sep=-\pgflinewidth, below of=A1]{2 & 5 & 8\\};
    \matrix[right=.5em of A2] (B2) [matrix of math nodes, nodes={MyStyle, anchor=center}, column sep=-\pgflinewidth]{1 & 4 & 7\\};
    \matrix (A3) [matrix of math nodes, nodes={MyStyle, anchor=center}, column sep=-\pgflinewidth, below of=A2]{|[draw,fill=red!30]|2 & |[draw,fill=blue!30]|5 & |[draw,fill=blue!30]|8\\};
    \matrix[right=.5em of A3] (B3) [matrix of math nodes, nodes={MyStyle, anchor=center}, column sep=-\pgflinewidth]{|[draw,fill=red!30]|1 & |[draw,fill=red!30]|4 & |[draw,fill=blue!30]|7\\};
    \matrix (A4) [matrix of math nodes, nodes={MyStyle, anchor=center}, column sep=-\pgflinewidth, below of=A3]{1 & 2 & 4\\};
    \matrix[right=.5em of A4] (B4) [matrix of math nodes, nodes={MyStyle, anchor=center}, column sep=-\pgflinewidth]{5 & 7 & 8\\};
    \begin{scope}[->,>=stealth]
        \draw (A1-1-2.south) -- (A2-1-2.north);
        \draw (B1-1-2.south) -- (B2-1-2.north);
        \node[align=left, anchor=east] (a) at (A2-1-1.west) {Sorting each block};
        \node[align=left, anchor=east] (a) at (A3-1-1.west) {Partitioning};
        \draw[red] (A3-1-1.south) -- (A4-1-2.north);
        \draw[red] (B3-1-1.south east) -- (A4-1-2.north);
        \draw[blue] (A3-1-3.south west) -- (B4-1-2.north);
        \draw[blue] (B3-1-3.south) -- (B4-1-2.north);
        \node[align=left, anchor=east] (a) at (A4-1-1.west) {Merging of partitions};
    \end{scope}
    \end{tikzpicture}}
    \caption{Example of a parallel sort based on samplesort. First, it divides an input sequence into two blocks and sorts each block. Next, it partitions each sorted block, where 4 is selected as the pivot in this example. Finally, it collects the partitions in each block and merges them.}
    \label{fig:parallelSort}
\end{figure}
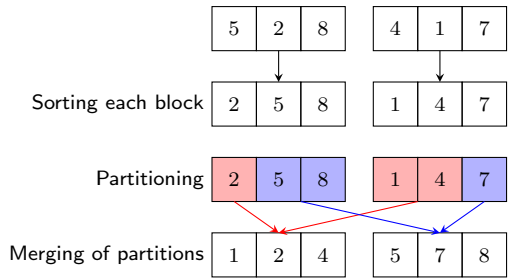
\figref{fig:parallelSort} shows an example for the case $A=\qty{5,2,8,4,1,7}$, $n_\text{B}=2$, and $n_\text{P}=2$. We first divide $A$ into two blocks of three elements and sort each block (step~\ref{procedure:blockSort}). Assuming that 4 is selected as the pivot $P_1$ (step~\ref{procedure:pivotSelection}), we rearrange each block into two partitions, less than or equal to 4 and greater than 4 (step~\ref{procedure:partitioning}), merge the partitions, and put them in order (step~\ref{procedure:merge}). As a result, the input sequence is sorted as $\qty{1,2,4,5,7,8}$.

Since steps~\ref{procedure:blockSort} and~\ref{procedure:merge} are the most time-consuming parts, optimizing these parts is vital to speed up samplesort. We compare variants of quicksort as sequential sorting, implement a few multiway merging algorithms and explore the fastest combinations. In addition, the parallel efficiency of samplesort depends significantly on how to select pivots. We also investigate how different pivots selection methods affect parallel efficiency, mainly when many elements are duplicated.

\subsection{Sequential sorting}\label{sec:quicksort}
Quicksort~\cite{quicksort} is an algorithm that sorts an entire sequence by recursively repeating partitioning. Partitioning is selecting a pivot element $P$ from an input sequence $A_0,\dots,A_{N-1}$, where $N$ is the number of elements, and rearranging the elements so that $A_0,\dots,A_k\leq P$ and $A_k,\dots,A_{N-1}\geq P$ hold for $k\in\qty{0,\dots,N-1}$. The average time complexity of quicksort is $O\qty(N\log N)$.

Supposing the maximum or minimum element of the input sequence is selected as the pivot for every partitioning, the time complexity of quicksort is worst and $O\qty(N^2)$. Therefore introsort~\cite{introsort} is often used to improve the worst-case time complexity of quicksort to $O\qty(N\log N)$. Introsort is an algorithm that switches to heapsort from quicksort when the depth of the recursive call of quicksort reaches a certain depth limit. If the depth limit is $O\qty(\log N)$, then the worst-case time complexity of introsort is $O\qty(N\log N)$.

To take advantage of the instruction pipeline of a processor, even when a program contains conditional branches, processors try to predict one of the branches and speculatively execute it. However, the pipelines of modern processors are long, and poor prediction reduces the program's performance. BlockQuicksort~\cite{blockQsort} uses block partitioning to prevent reducing the performance of quicksort due to branch mispredictions. The original partitioning~\cite{quicksort} scans the input sequence from both ends while comparing and exchanging the elements with the pivot. On the other hand, block partitioning stores the positions of elements to be exchanged in a buffer and exchanges them after the scanning is completed. In modern CPUs, comparing an element to a pivot and storing its position can be performed without conditional branches. For example, the CSET or CINC instruction in ARMv8 processor can perform it, reducing the number of conditional branches in partitioning.

In pattern-defeating quicksort~\cite{pdqsort}, the input sequence is partitioned into three parts: less than the pivot, equal to the pivot, and greater than the pivot. When the number of distinct elements $k$ is sufficiently small, the sorting time is $O\qty(Nk)$. A partitioning with a significant size bias can reduce the performance, called a bad partition. Pattern-defeating quicksort stops the recursive call when the number of bad partitions reaches $\lfloor\log N\rfloor$, enabling a switch to heapsort more precisely than introsort.

\subsection{Merging}\label{sec:merging}
We can efficiently merge $n_\text{B}$ sorted sequences using data structures. By organizing the head of each sorted sequence in the binary heap or the selection tree~\cite{Knuth1998} and retrieving the elements in ascending order, we can obtain a single sorted sequence. Since the calculation cost is $O\qty(\log n_\text{B})$ for retrieving the smallest element and updating the data structure, the total time for merging is $O\qty(n_\text{M}\log n_\text{B})$, where $n_\text{M}$ is the total number of elements to be merged in a partition.

Sequential sorting enables multiway merging without data structures. We can gather $n_\text{B}$ sorted sequences into a single sorted sequence by copying them into an output array and sorting it. Although the time complexity of this method is $O\qty(n_\text{M}\log n_\text{M})$, the cache efficiency should be higher because of the locality of memory accesses.

\subsection{Pivots selection and partitioning}\label{sec:pivots_selection}
The parallel efficiency of samplesort is decreased unless the pivots are selected so that the total number of elements to be merged is equally balanced between threads. Parallel Sorting by Regular Sampling~(PSRS)~\cite{Shi1992} samples $n_\text{P}-1$ elements from each block, sorts them and selects $n_\text{P}-1$ pivots from the $n_{\rm B}(n_\text{P}-1)$ samples. It picks the samples and the pivots from sorted sequences at regular intervals so that the total number of elements to be merged is as equal as possible. The time complexity of the selection of the pivots is $O\qty(n_\text{B}n_\text{P}\log\qty(n_\text{B}n_\text{P}))$, corresponding to the sequential sorting of all the samples.

\textsc{sortlib}~\footnote{https://github.com/jmakino/sortlib}, which is a reasonably fast samplesort implementation for A64FX and other architectures, randomly samples elements from each block without sorting the block, sorts all the samples, and selects pivots at regular intervals. Since the total number of the samples is proportional to the number of elements $N$ and inversely proportional to the number of threads $t$, the time complexity of the selection of the pivots is $O\qty(\qty(N/t)\log\qty(N/t))$.

\begin{figure}
    \begin{subfigure}[t]{\linewidth}
        \centering
        {\sffamily
        \begin{tikzpicture}[MyStyle/.style={draw, minimum width=2em, minimum height=2em, outer sep=0pt}]
        \matrix (A1) [matrix of math nodes, nodes={MyStyle, anchor=center}, column sep=-\pgflinewidth]{|[draw,fill=red!30]|1 & |[draw,fill=red!30]|1 & |[draw,fill=blue!30]|4\\};
        \matrix[right=.5em of A1] (B1) [matrix of math nodes, nodes={MyStyle, anchor=center}, column sep=-\pgflinewidth]{|[draw,fill=red!30]|1 & |[draw,fill=red!30]|1 & |[draw,fill=red!30]|1\\};
        \matrix[right=2em of A1] (A2) [matrix of math nodes, nodes={MyStyle, anchor=center}, column sep=-\pgflinewidth, below of=A1]{1 & 1 & 1 & 1 & 1\\};
        \matrix[right=.5em of A2] (B2) [matrix of math nodes, nodes={MyStyle, anchor=center}, column sep=-\pgflinewidth]{4\\};
        \begin{scope}[->,>=stealth]
            \draw[red] (A1-1-1.south east) -- (A2-1-3.north);
            \draw[red] (B1-1-2.south) -- (A2-1-3.north);
            \draw[blue] (A1-1-3.south) -- (B2-1-1.north);
        \end{scope}
        \end{tikzpicture}}
        \caption{Partitioning with the pivot \(P_1=1\). Since more elements are in the first partition than in the second, the merging time is highly imbalanced between partitions.}
        \label{subfig:psrs}
    \end{subfigure}
    \begin{subfigure}[t]{\linewidth}
        \centering
        {\sffamily
        \begin{tikzpicture}[MyStyle/.style={draw, minimum width=2em, minimum height=2em, outer sep=0pt}]
        \matrix (A3) [matrix of math nodes, nodes={MyStyle, anchor=center}, column sep=-\pgflinewidth, below of=A2]{|[draw,fill=red!30]|1 & |[draw,fill=red!30]|1 & |[draw,fill=blue!30]|4\\};
        \matrix[right=.5em of A3] (B3) [matrix of math nodes, nodes={MyStyle, anchor=center}, column sep=-\pgflinewidth]{|[draw,fill=red!30]|1 & |[draw,fill=blue!30]|1 & |[draw,fill=blue!30]|1\\};
        \matrix (A4) [matrix of math nodes, nodes={MyStyle, anchor=center}, column sep=-\pgflinewidth, below of=A3]{1 & 1 & 1\\};
        \matrix[right=.5em of A4] (B4) [matrix of math nodes, nodes={MyStyle, anchor=center}, column sep=-\pgflinewidth]{1 & 1 & 4\\};
        \begin{scope}[->,>=stealth]
            \draw[red] (A3-1-1.south east) -- (A4-1-2.north);
            \draw[red] (B3-1-1.south) -- (A4-1-2.north);
            \draw[blue] (A3-1-3.south) -- (B4-1-2.north);
            \draw[blue] (B3-1-3.south west) -- (B4-1-2.north);
        \end{scope}
        \end{tikzpicture}}
        \caption{Partitioning in PSES with \(P_1=1\) and \(c_1=3\). The first partition contains elements less than one and three 1s; the second contains the remaining two 1s and elements greater than 1. Since both partitions have the same number of elements, the merging time is balanced between partitions.}
        \label{subfig:pses}
    \end{subfigure}
    \caption{Partitioning when there are few distinct elements.}
    \label{fig:dup}
\end{figure}
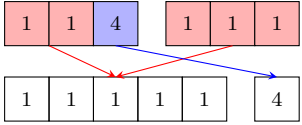
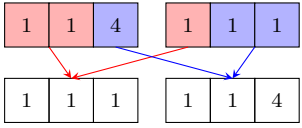
When the value of a large number of elements is duplicated, partitioning with only pivots may cause the total number of elements to be highly imbalanced between partitions. For example, in the case of input sequence $A=\qty{1,1,4,1,1,1}$, whatever the pivots are selected, the maximum number of total elements in a partition is more than or equal to 5 in PSRS (\figref{subfig:psrs}). Supposing the total number of elements in each partition is imbalanced, the time required for each merging is highly imbalanced, and parallel efficiency decreases as the number of threads increases. Parallel Sorting using Exact Splitting~(PSES)~\cite{Siebert2011} can overcome this issue. It selects pivots $P_k$ satisfying Equation~\eqref{eq:pses_condition}, and calculates the number $c_k$ defined in Equation~\eqref{eq:pses_count}.
\begin{align}
    \label{eq:pses_condition}
    \abs{\qty{x\in A \mid x < P_k }}\leq k\frac{N}{n_\text{P}}\leq \abs{\qty{x\in A \mid x \leq P_k }}
\end{align}
\begin{align}
    \label{eq:pses_count}
    c_k = k\frac{N}{n_\text{P}} - \abs{\qty{x\in A \mid x < P_k }}
\end{align}
The $k$th partition $\qty(0\leq k<n_\text{P})$ contains all elements $x$ satisfying $P_k<x< P_{k+1}$ and $c_k$ elements equal to $P_{k+1}$. This method equalizes the total number of elements in each partition, no matter how few distinct elements exist. After counting the elements that are less than $P_k$ and equal to $P_k$ in each block using a binary search, it checks whether Equation~\eqref{eq:pses_condition} is satisfied. Thus $P_k$ and $c_k$ can be determined in $O\qty(n_\text{B}\log N\log \lceil N/n_\text{B}\rceil)$ time. In the example where $A=\qty{1,1,4,1,1,1}$ and $n_\text{P}=2$, there are no elements less than 1 and five elements less than or equal to 1. By setting $P_1$ to 1 and $c_1$ to 3, the total number of elements in each partition becomes equal, as shown in \figref{subfig:pses}, allowing the merging cost for each partition to be balanced.

\section{Performance evaluations}\label{sec:evaluations}
We experimentally demonstrate that our parallel sorting programs are significantly more efficient than a commonly used one on the supercomputer Fugaku.
\subsection{Experimental settings}
We used the supercomputer Fugaku at the RIKEN Center for Computational Science (R-CCS) to evaluate the performance of our parallel sorting implementation. A node of Fugaku has one A64FX processor and 32 GB memory. The processor consists of four Core Memory Groups~(CMGs) and runs at 2.00 GHz. Each CMG has 12 computational cores, so A64FX has 48 computational cores in total.

We implemented PSRS and PSES as parallel sorting algorithms. The number of blocks $n_\text{B}$ and partitions $n_\text{P}$ were the same as the number of threads used for the performance evaluation. PSES improves the parallel efficiency of the multiway merging for input arrays containing many duplicate elements, but the cost of selecting pivots is higher than that of PSRS. Accordingly, we quantitatively compared the performance differences between them for inputs with many and few duplicate elements. We also implemented the selection tree as a data structure for multiway merging and compared it to the binary heap, which is a more common data structure used in~\cite{Siebert2011}.

The program is written in C++ with OpenMP thread parallelization and compiled by the Fujitsu compiler FCC 4.9.0 (clang mode) with \verb|-Ofast| optimization option. We used the \verb|numactl| command to assign a single computational core in each thread. Here, we bound the first 12 threads to the computational cores in CMG0 and the next 12 to the computational cores in CMG1. In the same way, we assigned the remaining 24 threads to CMG2 and CMG3. Input and output arrays were divided into blocks of the number of threads, and each block was allocated in the CMG closest to the thread that owned it.

\begin{table*}[t]
  \centering
    \caption[Data type and data size of one element in input sequences.]{Data type and data size of one element in input sequences. We sort Pair and Particle by the key (uint64\_t).}
  \label{tab:input_type}
  \begin{tabular}{c|cc|l}
      Sequences & Type & Size (Byte) & \\
    \hline
    \hline
       UniformInt & uint32\_t & 4 & uniform random 32 bits integers in $\qty[0, 2^{32}-1]$.\\
       UniformFloat & float & 4 & uniform random 32 bits floating point numbers in $[0, 1)$.\\
       AlmostSorted & uint32\_t & 4 & an almost sorted sequence consisting of 32 bits integers in $[0, N-1]$.\\
       Duplicate3 & uint32\_t & 4 & uniform random integers in $\qty{0, 1, 2}$.\\
       Pair & struct & 16 & key-index pairs. The keys are 64 bits uniform random integers in $\qty[0, 2^{64}-1]$.\\
       Particle & struct & 96 & key and particle data. The keys are 64 bits uniform random integers in $\qty[0, 2^{64}-1]$.\\
    \hline
  \end{tabular}
\end{table*}

We evaluated the performance of our sorting implementation for six input sequences shown in \tabref{tab:input_type} and the different numbers $N$ of the elements ($N=10^7$ and $10^8$). We constructed AlmostSorted data by swapping the positions of $\sqrt{N}$ elements chosen randomly from an increasing sequence from zero to $N-1$. Since the elements of Duplicate3 can take only three values, it contains many duplicate elements. Particle is a sequence of structures consisting of a key for sorting and data representing a particle in three-dimensional space (mass, position, velocity, acceleration, and potential. The total 88 bits in double precision), which are typical components of a particle structure in gravitational $N$-body simulations. This data type is used to evaluate the performance of sorting particles by some key. We measured the elapsed time for each sorting 20 times and used their average as the performance of our implementation.

\subsection{Comparison of parallel sorting algorithms}
\begin{figure*}[t]
    \begin{subfigure}[t]{\columnwidth}
        \centering
        \includegraphics[width=\linewidth,clip]{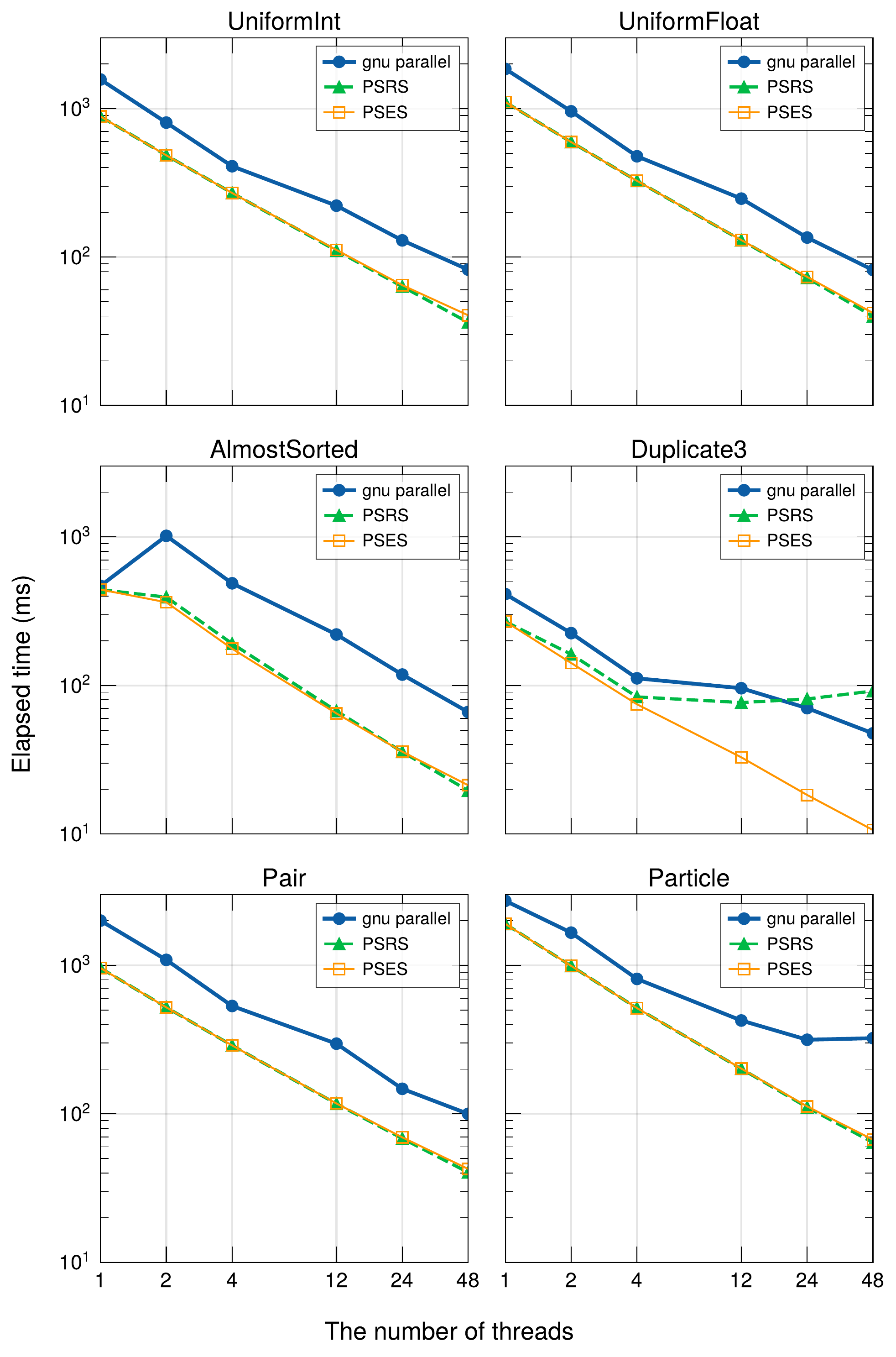}
        \caption{\(N=10^7\).}
        \label{subfig:parallel7}
    \end{subfigure}
    \begin{subfigure}[t]{\columnwidth}
        \centering
        \includegraphics[width=\linewidth,clip]{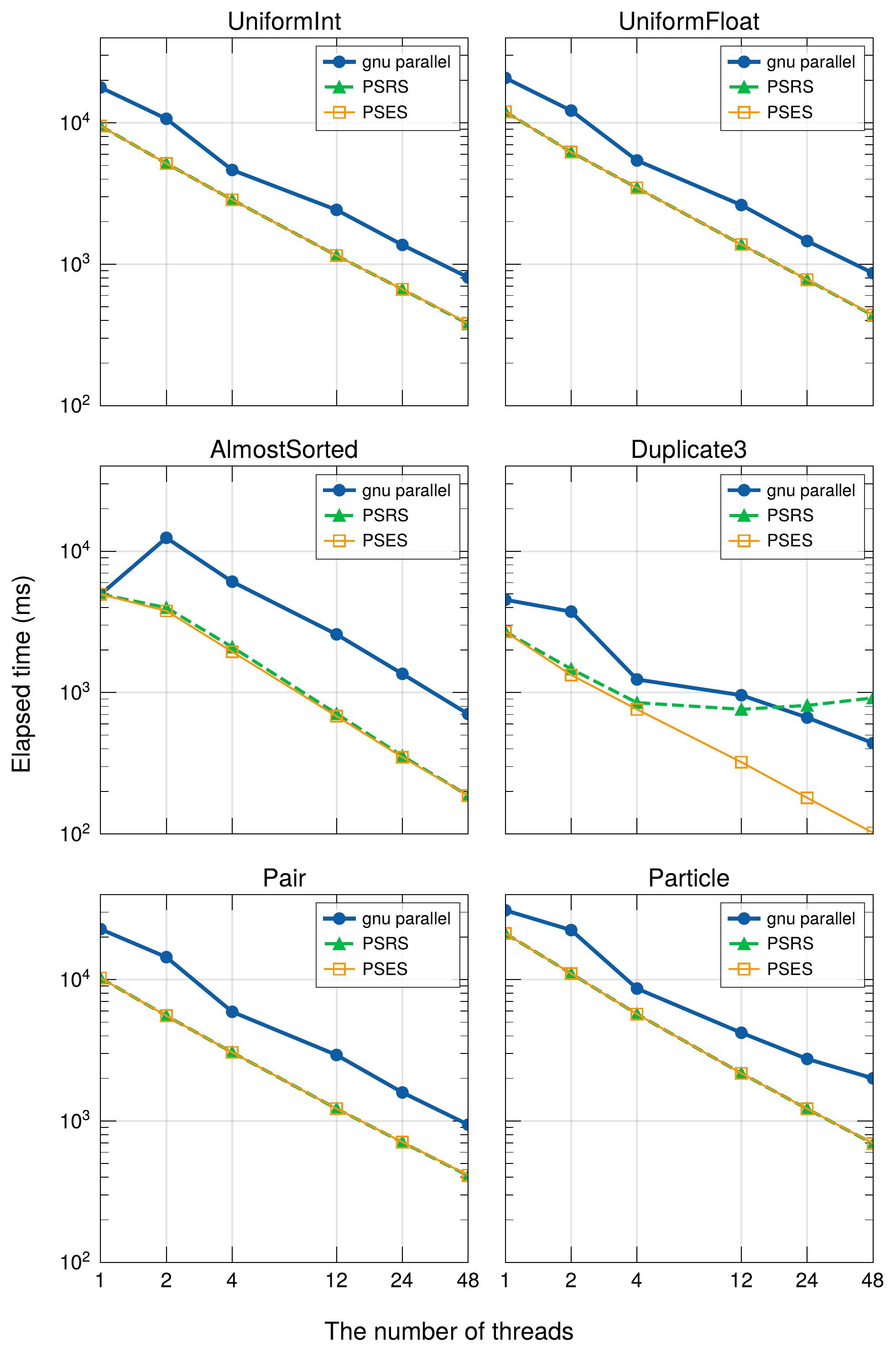}
        \caption{\(N=10^8\).}
        \label{subfig:parallel8}
    \end{subfigure}
    \caption{Elapsed time for parallel sorting algorithms. We compare our implementations of PSRS and PSES as well as \_\_gnu\_parallel::sort in libstdc++. We use BlockQuicksort for sequential sorting and our selection tree for multiway merging in both PSRS and PSES.}
    \label{fig:parallel}
\end{figure*}
\begin{figure*}[t]
    \begin{subfigure}[t]{\columnwidth}
        \centering
        \includegraphics[width=\linewidth,clip]{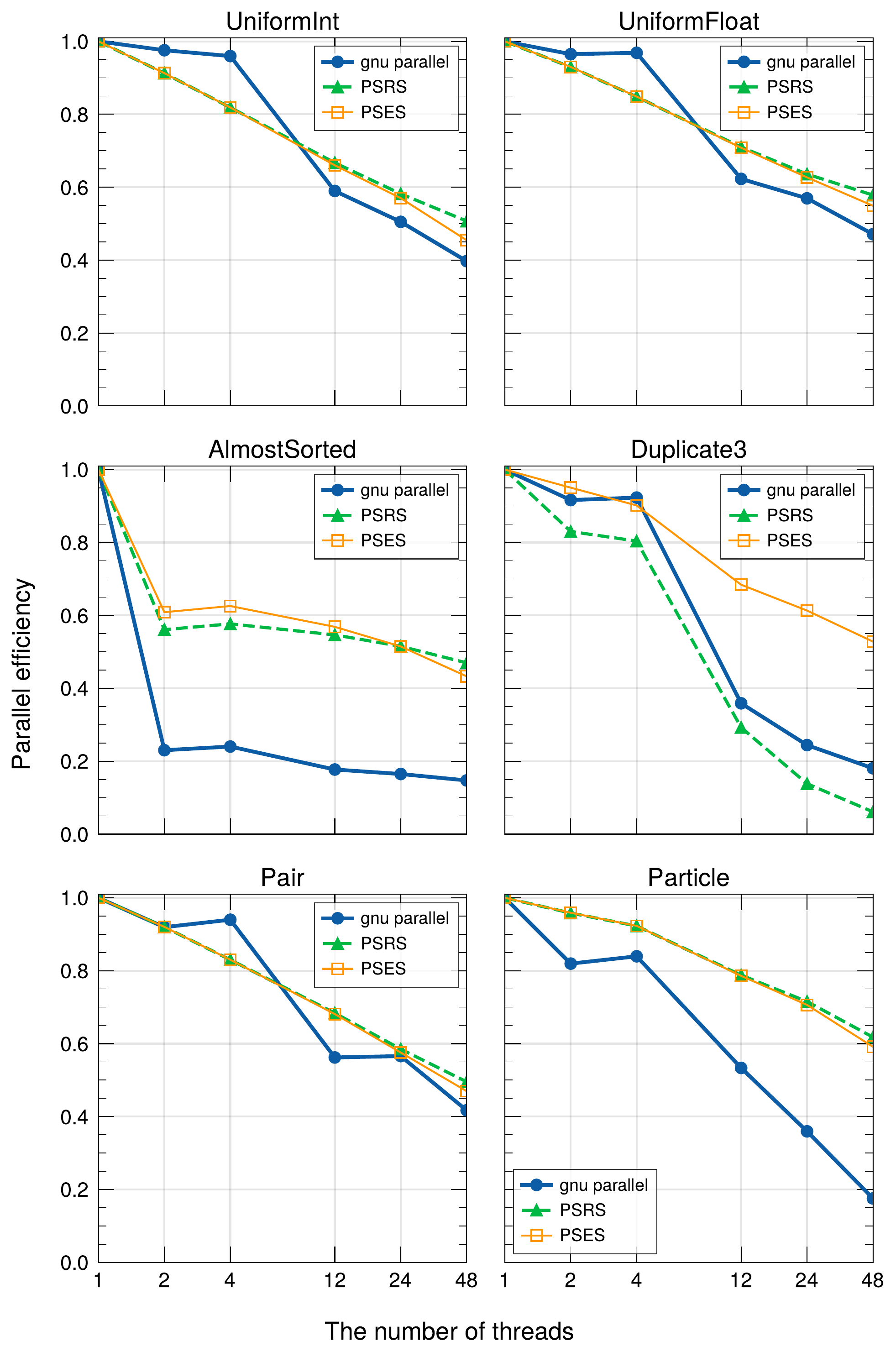}
        \caption{\(N=10^7\).}
        \label{subfig:efficiency7}
    \end{subfigure}
    \begin{subfigure}[t]{\columnwidth}
        \centering
        \includegraphics[width=\linewidth,clip]{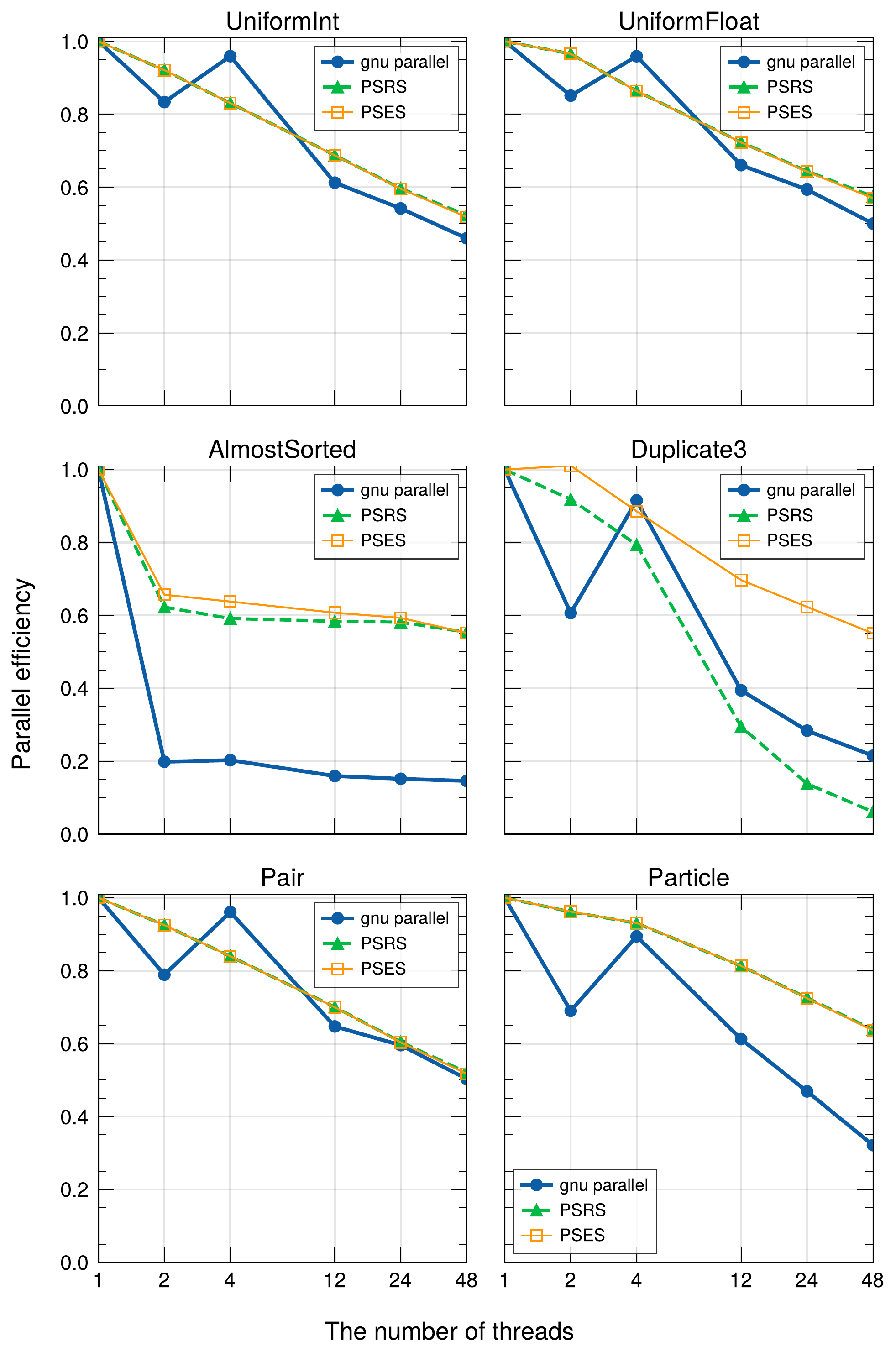}
        \caption{\(N=10^8\).}
        \label{subfig:efficiency8}
    \end{subfigure}
    \caption{Parallel efficiency of parallel sorting algorithms. We compare our implementations of PSRS and PSES as well as \_\_gnu\_parallel::sort in libstdc++. We use BlockQuicksort for sequential sorting and our selection tree for multiway merging in both PSRS and PSES.}
    \label{fig:efficiency}
\end{figure*}
\figref{fig:parallel} shows the elapsed time for parallel sorting algorithms. We use BlockQuicksort for sequential sorting and our selection tree for multiway merging in both PSRS and PSES. Here, \_\_gnu\_parallel::sort is a parallel multiway mergesort~\cite{MCSTL} implementation in the parallel mode~\cite{gnu_parallel_libstdc++} of the GNU C++ Standard Library called libstdc++. Its sorting procedure is similar to PSES, using std::sort for sequential sorting and the selection tree for $k$-way merging for $k$ greater than four. However, when there are many duplicate elements, it does not necessarily equalize the number of elements in each partition. This is due to additional constraints on the position of pivots to make the sorting stable.

For Particle with $N=10^7$, the time for sorting by \_\_gnu\_parallel::sort does not decrease as the number of threads increases above 24. In contrast, the sorting time for PSRS and PSES monotonically decreases to 48 threads. For all input sequences with $N=10^7$ and $10^8$ using 48 threads, PSES is more than twice as fast as \_\_gnu\_parallel::sort. The time for sorting by PSRS is almost the same as PSES for the input sequences with few duplicate elements. However, for Duplicate3, the elapsed time of PSRS does not decrease as the number of threads increases above four threads regardless of the number of elements, showing the stark difference from PSES. This saturation is because when the number of threads is larger than the number of distinct elements, no matter how we select the pivots, the number of elements to be merged by each thread becomes highly imbalanced, as described in Section~\ref{sec:pivots_selection}.

Those trends are further highlighted in \figref{fig:efficiency} that shows parallel efficiency. PSES has relatively excellent parallel efficiency for all input sequences above 12 threads, maintaining about 0.5 and 0.4 up to 48 threads with $N=10^8$ and $10^7$, respectively. PSRS shows nearly the same parallel efficiency as PSES except for Duplicate3, but its efficiency with 48 threads for Duplicate3 is the lowest and about a tenth of PSES. For the input sequences except for Duplicate3, \_\_gnu\_parallel::sort has the lowest parallel efficiency. It is prominently low for AlmostSorted, indicating that one-thread sequential sorting for AlmostSorted is faster than the other inputs.

All the algorithms have similar or higher parallel efficiency with $N=10^8$ than $N=10^7$ when the number of threads is large. In this study, we have parallelized the parts with relatively large time complexity for $N$, i.e., sorting each block and merging partitions. Therefore, in the case of increasing $N$ and keeping the number of threads, the time complexity becomes relatively more prominent on the parallel parts than on the non-parallel parts, resulting in higher parallel efficiency for larger $N$. In particular, for Pair and Particle, whose data size is larger than the others, the parallel efficiency would become higher with increasing $N$ because the data copy takes longer in the parallel parts.

\subsection{Comparison of sequential sorting algorithms}
\begin{figure}[t]
    \centering
    \includegraphics[width=\linewidth,clip]{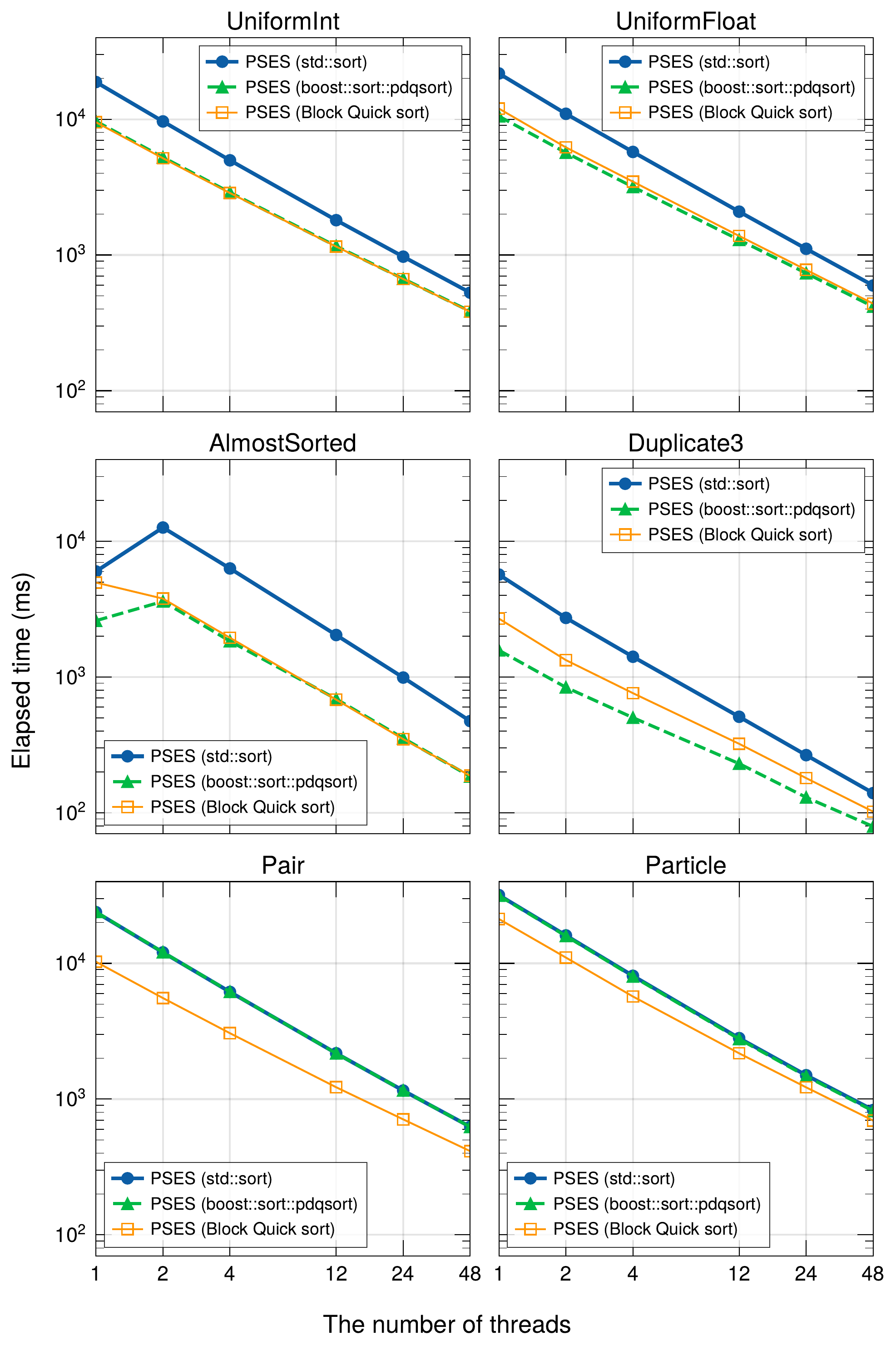}
    \caption{Elapsed time for PSES (\(N=10^8\)) with different block sorting algorithms. We compare std::sort in libstdc++, boost::sort::pdqsort in Boost Libraries, and BlockQuicksort. We use our selection tree implementation for multiway merging.}
    \label{fig:pses_seq8}
\end{figure}
\figref{fig:pses_seq8} shows the elapsed time for PSES using different algorithms to sort each block sequentially. We compare three algorithms: introsort, pattern-defeating quicksort, and BlockQuicksort, and use our selection tree for the multiway merging of partitions in all cases. The previous study~\cite{Siebert2011} argued that merge sort is more suitable than quicksort for sorting blocks because the execution time of quicksort varies depending on input sequences. However, the above three algorithms improve the worst-case time complexity of quicksort and are generally much faster than merge sort, so we compare them in this study. We use std::sort in libstdc++ as an implementation of introsort and boost::sort::pdqsort in Boost Libraries as an implementation of pattern-defeating quicksort. The publicly available code~\footnote{https://github.com/weissan/BlockQuicksort} of BlockQuicksort has several implementations with different ways of, e.g., selecting a pivot; we use one of these codes, blocked\_double\_pivot\_check\_mosqrt::sort.

BlockQuicksort is the fastest for the input sequences except for Duplicate3. Pattern-defeating quicksort is the fastest for Duplicate3 because it is designed to run faster when the number of distinct elements is small, as described in Section~\ref{sec:quicksort}. In most cases, std::sort does not run faster than the others.

\subsection{Comparison of multiway merging algorithms}
\begin{figure}[t]
    \centering
    \includegraphics[width=\linewidth,clip]{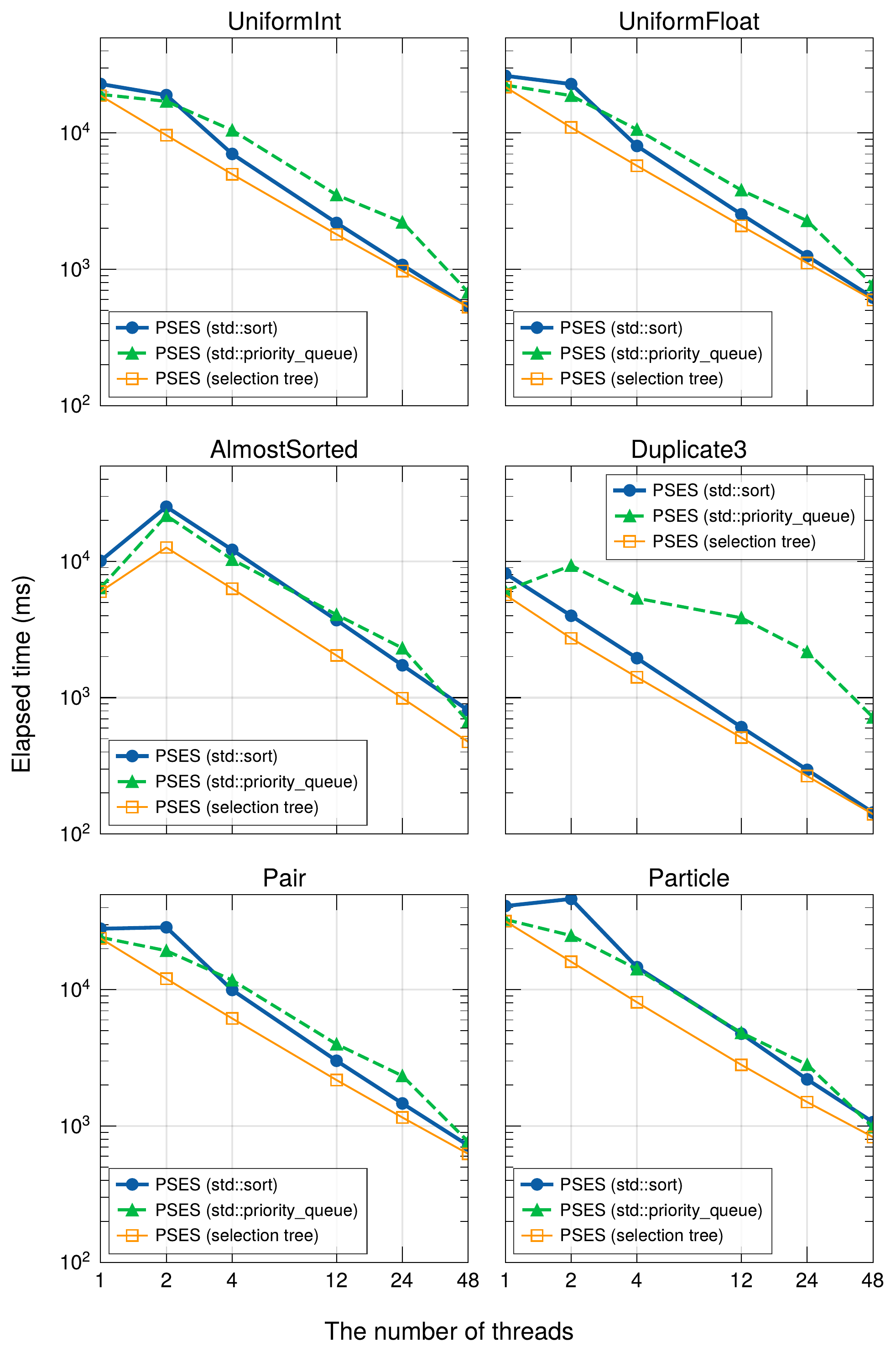}
    \caption{Elapsed time for PSES (\(N=10^8\)) with different algorithms for merging partitions. We compare std::priority\_queue in libstdc++ and our selection tree, as well as sequential sorting by std::sort without data structures. We use std::sort for sorting each block.}
    \label{fig:pses_merge8}
\end{figure}
\figref{fig:pses_merge8} shows the elapsed time for PSES using different multiway merging algorithms to merge partitions. We compare three algorithms: the binary heap, the selection tree, and sequential sorting without data structures. We use std::priority\_queue in libstdc++ as an implementation of the binary heap and std::sort for sequential sorting to merge partitions. In all cases, we use std::sort for sorting each block.

Our selection tree implementation is the fastest for all the input sequences. However, for 24 threads or more, std::sort shows similar performance with the selection tree for input sequences UniformInt, UniformFloat, and Duplicate3, even though this sort typically requires movements of elements more frequently than the selection tree. The reason for relatively good performance would be that this sort does not use external data structures and is cache efficient. On the other hand, this sort becomes slower for Pair and Particle because those data size is large, causing poor cache performance. With 12 threads or more, std::sort is more than 1.5 times slower for AlmostSorted than UniformInt, even though the selection tree gives similar results. Using a profiler provided by Fujitsu, we examined the differences in merging by std::sort for these two inputs and found that std::sort execution time increased significantly for AlmostSorted. Furthermore, std::sort finally called heap sort for AlmostSorted, but not for UniformInt. Thus, we consider that the order of the elements to be merged for AlmostSorted may degrade the performance of std::sort. The sorting time of std::priority\_queue raises even from two to four threads for Duplicate3, Pair, and Particle. In most cases, it is slower than using std::sort for multiway merging, indicating that containers in libstdc++ are not implemented to run efficiently on Fugaku.

\section{Conclusion}\label{sec:conclusion}
We have implemented two multi-threaded sorting algorithms that can be executed on modern supercomputers such as Fugaku. Both algorithms are based on samplesort but differ in selecting pivots. The first algorithm divides the input sequence into multiple blocks, then sorts each block and selects pivots by sampling from each block at regular intervals. The second algorithm uses the binary search to select pivots so that the number of elements in each partition is equal. Then we compared the performance of the two algorithms with different sequential sorting and multiway merging algorithms. With the combination of BlockQuicksort for sequential sorting and the selection tree for multiway merging, the second algorithm shows high speed and high parallel efficiency for various input data types and data sizes.

In future work, we will consider making the performance evaluation of parallel radix sort and in-place parallel sorting algorithms. Radix sort may be faster than comparison sorts. Parallel quicksort and In-place Parallel Super Scalar Samplesort (IPS$^4$o)~\cite{axtmann2020engineering} are representative in-place parallel sorting algorithms at present. IPS$^4$o is a novel parallel sorting algorithm that recursively performs multiway partitioning, improving Super Scalar Samplesort's method~\cite{S4o}. It is an in-place algorithm because the additional memory usage at each recursive level is independent of the length of the input sequence. Both parallel sorting programs implemented in this study assume that the input and output sequences are placed separately in memory. In other words, those require allocating additional memory for the output sequence with the same size as the input. It is essential to establish high-performance in-place parallel sorting programs that run on supercomputers with a small amount of memory per node, like Fugaku, or on systems with low Bytes-per-Flop~(B/F) ratios, where efficient use of cache is crucial.

\begin{acknowledgment}
This work has been supported by IAAR Research Support Program in Chiba University Japan, MEXT/JSPS KAKENHI (Grant Number JP21H01122), MEXT as ``Program for Promoting Researches on the Supercomputer Fugaku'' (JPMXP1020200109), and JICFuS.
\end{acknowledgment}

\bibliographystyle{ipsjsort-e}

\begin{thebibliography}{10}

\bibitem{axtmann2020engineering}
Axtmann, M., Witt, S., Ferizovic, D. and Sanders, P.: Engineering In-place
  (Shared-memory) Sorting Algorithms, Computing Research Repository (CoRR)
  (Sept. 2020).

\bibitem{blockQsort}
Edelkamp, S. and Wei\ss{}, A.: BlockQuicksort: Avoiding Branch Mispredictions
  in Quicksort, {\em ACM J. Exp. Algorithmics},  Vol.~24 (2019).

\bibitem{Samplesort}
Frazer, W.~D. and McKellar, A.~C.: Samplesort: A Sampling Approach to Minimal
  Storage Tree Sorting, {\em J. ACM},  Vol.~17, No.~3, p.\ 496–507 (online),
  \doi{10.1145/321592.321600} (1970).

\bibitem{quicksort}
Hoare, C. A.~R.: {Quicksort}, {\em The Computer Journal},  Vol.~5, No.~1, pp.\
  10--16 (1962).

\bibitem{Knuth1998}
Knuth, D.~E.: {\em The Art of Computer Programming, Volume 3: (2nd Ed.) Sorting
  and Searching}, Addison Wesley Longman Publishing Co., Inc., USA (1998).

\bibitem{introsort}
MUSSER, D.~R.: Introspective Sorting and Selection Algorithms, {\em Software:
  Practice and Experience},  Vol.~27, No.~8, pp.\ 983--993 (1997).

\bibitem{RegionsSort}
Obeya, O., Kahssay, E., Fan, E. and Shun, J.: Theoretically-Efficient and
  Practical Parallel In-Place Radix Sorting, {\em The 31st ACM Symposium on
  Parallelism in Algorithms and Architectures}, SPAA '19, New York, NY, USA,
  Association for Computing Machinery, p.\ 213–224 (online),
  \doi{10.1145/3323165.3323198} (2019).

\bibitem{pdqsort}
Peters, O. R.~L.: Pattern-defeating Quicksort, {\em CoRR},  Vol.~abs/2106.05123
  (2021).

\bibitem{S4o}
Sanders, P. and Winkel, S.: Super Scalar Sample Sort, {\em Algorithms -- ESA
  2004} (Albers, S. and Radzik, T., eds.), Berlin, Heidelberg, Springer Berlin
  Heidelberg, pp.\ 784--796 (2004).

\bibitem{Shi1992}
Shi, H. and Schaeffer, J.: Parallel sorting by regular sampling, {\em Journal
  of Parallel and Distributed Computing},  Vol.~14, No.~4, pp.\ 361--372
  (1992).

\bibitem{PBBS}
Shun, J., Blelloch, G.~E., Fineman, J.~T., Gibbons, P.~B., Kyrola, A.,
  Simhadri, H.~V. and Tangwongsan, K.: Brief Announcement: The Problem Based
  Benchmark Suite, {\em Proceedings of the Twenty-Fourth Annual ACM Symposium
  on Parallelism in Algorithms and Architectures}, SPAA '12, New York, NY, USA,
  Association for Computing Machinery, p.\ 68–70 (online),
  \doi{10.1145/2312005.2312018} (2012).

\bibitem{Siebert2011}
Siebert, C. and Wolf, F. G.~E.: {A} scalable parallel sorting algorithm using
  exact splitting, Technical report, Aachen (2011).

\bibitem{gnu_parallel_libstdc++}
Singler, J. and Konsik, B.: The GNU Libstdc++ Parallel Mode: Software
  Engineering Considerations, {\em Proceedings of the 1st International
  Workshop on Multicore Software Engineering}, IWMSE '08, New York, NY, USA,
  Association for Computing Machinery, p.\ 15–22 (2008).

\bibitem{MCSTL}
Singler, J., Sanders, P. and Putze, F.: MCSTL: The Multi-Core Standard Template
  Library, {\em Proceedings of the 13th International Euro-Par Conference on
  Parallel Processing}, Euro-Par'07, Berlin, Heidelberg, Springer-Verlag, p.\
  682–694 (2007).

\bibitem{Solomonik2010}
Solomonik, E. and Kalé, L.~V.: Highly scalable parallel sorting, {\em 2010
  IEEE International Symposium on Parallel and Distributed Processing (IPDPS)},
  pp.\ 1--12 (online), \doi{10.1109/IPDPS.2010.5470406} (2010).

\bibitem{Pquick}
Tsigas, P. and Zhang, Y.: A simple, fast parallel implementation of Quicksort
  and its performance evaluation on SUN Enterprise 10000, {\em Eleventh
  Euromicro Conference on Parallel, Distributed and Network-Based Processing,
  2003. Proceedings.}, pp.\ 372--381 (2003).

\end{thebibliography}

\end{document}